\documentclass[runningheads]{llncs}

\usepackage{graphicx}
\usepackage{booktabs}
\usepackage{multirow}
\usepackage{rotating}
\usepackage{breakcites}

\usepackage[colorinlistoftodos]{todonotes}

\usepackage[resetlabels,labeled]{multibib}
\newcites{SR}{References included in the systematic literature overview}

\begin{document}
\title{A Systematic Literature Review of Empirical Research on Quality Requirements\thanks{The work is partially supported by a research grant for the ORION project (reference number 20140218) from The Knowledge Foundation in Sweden.}}
\titlerunning{Systematic literature review on QRs}
\author{Thomas Olsson\inst{1}\orcidID{0000-0002-2933-1925} \and
S\'everine Sentilles\inst{2}\orcidID{0000-0003-0165-3743} \and
Efi Papatheocharous\inst{1}\orcidID{0000-0002-5157-8131}}
\authorrunning{T. Olsson et al.}

\institute{RISE Research Institutes of Sweden AB, Sweden 
\email{\{thomas.olsson,efi.papatheocharous\}@ri.se}\\
\url{https://ri.se}
\and
M\"alardalen University, Sweden\\
\url{https://www.mdh.se} \\
\email{severine.sentilles@mdh.se}}

\maketitle              
\begin{abstract}
Quality requirements deal with how well a product should perform the intended functionality, such as start-up time and learnability. Researchers argue they are important and at the same time studies indicate there are deficiencies in practice. 

Our goal is to review the state of evidence for quality requirements. We want to understand the empirical research on quality requirements topics as well as evaluations of quality requirements solutions. 

We used a hybrid method for our systematic literature review. We defined a start set based on two literature reviews combined with a keyword-based search from selected publication venues. We snowballed based on the start set. 

We screened 530 papers and included 84 papers in our review. Case study method is the most common (43), followed by surveys (15) and tests (13). We found no replication studies. The two most commonly studied themes are 1) Differentiating characteristics of quality requirements compared to other types of requirements, 2) the importance and prevalence of quality requirements. Quality models, QUPER, and the NFR method are evaluated in several studies, with positive indications. Goal modeling is the only modeling approach evaluated. However, all studies are small scale and long-term costs and impact are not studied. 

We conclude that more research is needed as empirical research on quality requirements is not increasing at the same rate as software engineering research in general. We see a gap between research and practice. The solutions proposed are usually evaluated in an academic context and surveys on quality requirements in industry indicate unsystematic handling of quality requirements.

\keywords{Systematic literature review  \and Quality requirements \and Non-functional requirements \and Empirical evidence}
\end{abstract}
%

\section{Introduction}
Quality requirements -- also known as non-functional requirements -- are requirements related to how well a product or service is supposed to perform the intended functionality~\cite{glinz2007non}. Examples are start-up time, access control, and learnability~\cite{ISO25010}. Researchers have long argued the importance of quality requirements~\cite{ebert1998putting,lawrence2001top,paech2004non}. However, to what extent have problems and challenges with quality requirements been studied empirically? A recent systematic mapping study identified quality requirements as one of the emergent areas of empirical research~\cite{ambreen2018empirical}. There are several proposals over the years for how to deal with quality requirements, e.g., the NFR method~\cite{mylopoulos1992representing}, QUPER~\cite{regnell2008supporting}, quality models~\cite{dorr2003eliciting}, and i*~\cite{yu1997a}. However, to what extent have they been empirically validated? We present a systematic literature review of empirical studies on problems and challenges as well as validated techniques and methods for quality requirements engineering. 

Ambreen et al. conducted a systematic mapping study on empirical research in requirements engineering~\cite{ambreen2018empirical}, published in 2018. They found 270 primary studies where 36 papers were categorized as research on quality requirements. They concluded that empirical research on quality requirements is an emerging area within requirements engineering. Berntsson Svensson et al. carried out a systematic mapping study on empirical studies on quality requirements~\cite{BerntssonSvensson2010261}, published in 2010. They found 18 primary empirical studies on quality requirements. They concluded that there is a lack of unified view and reliable empirical evidence, for example, through replications and that there is a lack of empirical work on prioritization in particular. In our study, we follow up on the systematic mapping study om Ambreen et al.~\cite{ambreen2018empirical} by performing a systematic literature review in one of the highlighted areas. Our study complements Berntsson Svensson et al. study from 2010 by performing a similar systematic literature review 10 years later and by methodologically also using a snowball approach. 

There exist several definitions of quality requirements as well as names~\cite{glinz2007non}. Glinz defines a non-functional requirement as an attribute (such as performance or security) or a constraint on the system. The two prevalent terms are quality requirements and non-functional requirements. Both are used roughly as much and usually mean approximately the same thing. The ISO25010 defines quality in use as to whether the solution fulfills the goals with effectiveness, efficiency, freedom from risk, and satisfaction~\cite{ISO25010}. Eckhardt et al. analyzed 530 quality requirements and found that they described a behavior -- essentially a function~\cite{eckhardt2016non}. Hence, the term non-functional might be counter-intuitive. We use the term quality requirements in this paper. In layman's terms, we mean a quality requirement expresses \textit{how well} a solution should execute an intended function, as opposed to functional requirements which express \textit{what} the solution should perform. Furthermore, conceptually, we use the definition from Glinz~\cite{glinz2007non} and the sub characteristics of ISO25010 as the main refinement of quality requirements~\cite{ISO25010}. 

We want to understand from primary studies 1) what are the problems and challenges with quality requirements as identified through empirical studies, and 2) which quality requirements solutions have been empirically validated. We are motivated by addressing problems with quality requirements in practice and understanding why quality requirements is still, after decades of research, often reported as a troublesome area of software engineering in practice. Hence, we study which are the direct observations and experience with quality requirements. We define the following research questions for our systematic literature review: 
\begin{enumerate}
    \item[RQ1] Which empirical methods are used to study quality requirements? 
    \item[RQ2] What are the problems and challenges for quality requirements identified by empirical studies?
    \item[RQ3] Which quality requirements solution proposals have been empirically validated?
\end{enumerate}

We study quality requirements in general and therefore exclude papers focusing on specific aspects, e.g., on safety or user experience. 

We summarize the related literature reviews in Section~\ref{sec:RW}. We describe the hybrid method we used for our systematic literature review in Section~\ref{sec:Design}. Section~\ref{sec:results} elaborates on the findings from screening of 530 papers to finally include 84 papers from the years 1995 to 2019. We discuss the results and threats to validity in Section~\ref{sec:discussion}; empirical studies on quality requirements are -- in relative terms -- less common than other types of requirements engineering papers, there is a lack of longitudinal studies of quality requirements topics, we found very few replications. We conclude the paper in Section~\ref{sec:conclusion} with a reflection that there seems to be a divide between solutions proposed in an academic setting and the challenges and needs of practitioners. 

\section{Related work}\label{sec:RW}

A recent systematic mapping study on empirical studies on requirements engineering states that quality requirements are ``by far the most active among these emerging research areas"~\cite{ambreen2018empirical}. They classified 36 papers of the 270 they included as papers in the quality requirements area. In their mapping, they identify security and usability as the most common topics. These results are similar to that of Ouhbi et al. systematic mapping study from 2013~\cite{ouhbi2013software}. However, they had slightly different keywords in their search, including also studies on quality in the requirements engineering area, which is not necessarily the same as quality requirements. A systematic mapping study is suggested for a broader area whereas a systematic literature review for a narrower area which is studies in more depth~\cite{Kitchenham07guidelinesfor}. To our knowledge, there are no recent systematic literature reviews on quality requirements. 

Berntsson Svensson et al. performed a systematic literature review on empirical studies on managing quality requirements in 2010~\cite{BerntssonSvensson2010261}. They identified 18 primary studies. They classified 12 out of the 18 primary studies as case studies, three as experiments, two as surveys, and one as a mix of survey and experiment. They classified only four of the 18 studies as properly handling validity threats systematically. Their results indicate that there is a lack of replications and multiple studies on the same or similar phenomena. However, they identify a dichotomy between two views; those who argue that quality requirements need special treatment and others who argue quality requirements need to be handled at the same time as other requirements. Furthermore, they identify a lack of studies on prioritization of quality requirements. Bentsson Svensson et al. limited their systematic literature review to studies containing the keyword "software", whereas we did not in our study. Furthermore, Berntsson Svensson et al. performed a keyword-based literature search with a number of keywords required to be present in the search set. We used a hybrid approach and relied on snowballing instead of strict keywords. Lastly, we used Ivarsson and Gorschek~\cite{ivarsson2011method} for rigor, which entailed stricter inclusion criteria, i.e. as a result we did not include all studies from Berntsson Svensson et al. This, in combination with performing the study 10 years afterward, means we complement Berntsson Svensson both in terms of the method as well as studied period.

Alsaqaf et al. could not find any empirical studies on quality requirements in their 2017 systematic literature review on quality requirements in large-scale agile projects~\cite{alsaqaf2017quality}. They included studies on agile practices and requirements in general. Hence, their scope does not overlap significantly with ours. They found, however, 12 challenges to quality requirements in an agile context. For example, a focus on delivering functionality at the expense of architecture flexibility, difficulties in documenting quality requirements in user stories, and late validation of quality requirements. We do not explicitly focus on agile practices. Hence, there is a small overlap between their study and ours. 


\section{Design}\label{sec:Design}
\begin{figure}[bt]
\centering
\includegraphics[width=0.95\textwidth]{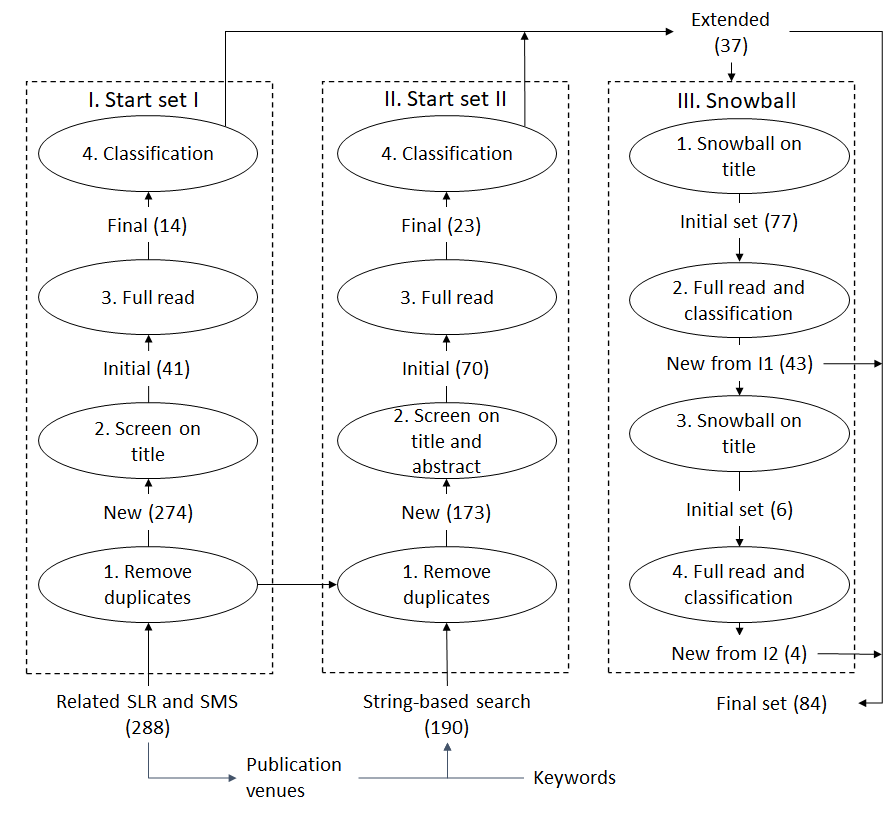} 
 \caption{We used two different approaches to create the start sets; Start set I is based on two other literature reviews, Start set II is created through a keyword-based search in relevant publication venues. The two start sets are combined and snowballed on to arrive at the final set of included papers. The numbers between the steps in each set are the number of references within that set. The numbers between the sets are the total number of references included in the final set.}
  \label{fig:process}
\end{figure}

We designed a systematic literature review using a hybrid method~\cite{mourao2017investigating}. The hybrid method combines a keyword-based search, typical of a systematic literature review~\cite{Kitchenham07guidelinesfor}, to define a start set and a snowball method~\cite{wohlin2014guidelines} to systematically find relevant papers. We base our study on two literature reviews~\cite{BerntssonSvensson2010261,ambreen2018empirical}, which we complement in a systematic way. The overall process is found in Figure~\ref{fig:process}. 

\subsection{Approach}
We decided to use a hybrid approach for our literature review~\cite{mourao2017investigating}. A standard keyword-based systematic literature review~\cite{Kitchenham07guidelinesfor} can result in a very large set of papers to review if keywords are not restrictive. On the other hand, having too restrictive keywords can result in a too-small set of papers. A snowball approach~\cite{wohlin2014guidelines}, on the other hand, is sensitive to the start set. If the studies are published in different communities not referencing each other, there is a risk of not finding relevant papers if the start set is limited to one community. Hence, we used a hybrid method where we combine the results from a systematic mapping study and a systematic literature review to give us one start set with a keyword-based search in the publication venues of the papers from the two review papers. 

\subsubsection{I. Start set I}
We defined Start set I for our systematic literature review by using a systematic mapping study on empirical evidence for requirements engineering in general~\cite{ambreen2018empirical} from 2018 and a systematic literature review from 2010~\cite{BerntssonSvensson2010261} with similar research questions as in our paper. 

The systematic literature review from 2010 by Berntsson Svensson et al. include 18 primary studies~\cite{BerntssonSvensson2010261}. However, we have different inclusion criteria (see Section~\ref{sec:Planning}). Hence, not all the references are included. In our final set, we included 10 of the 18 studies. 

The systematic mapping by Ambreem et al. from 2018 looks at empirical evidence in general for requirements engineering~\cite{ambreen2018empirical}. They included 270 primary studies. However, there are some duplicates in their list. They classified 36 papers to be in the quality requirements area. However, there is an overlap with the Berntsson Svensson et al. review~\cite{BerntssonSvensson2010261}. When we remove the already included papers from Berntsson Svensson et al., we reviewed 24 from Ambreem et al. and in the end included 4 of them.  

\subsubsection{II. Start set II}
To complement start set I, we also performed a keyword-based search. We have slightly different research questions than the two papers in the Start set I. Therefore, our search string is slightly different than that of Ambreen et al. and Berntsson Svensson et al. Also, the most recent references in Start set I are from 2014, i.e. five years before we performed our search. Hence, we also fill the gap of the papers published since 2014. We include all studies, not just studies from 2014 and onward, as our method and research questions are slightly different. 

We used the most frequent and highly ranked publication venues from Start set I to limit our search but still have a relevant scope. Table~\ref{tab:conf_journal} summarizes the included conferences and journals. Even though the publication venues included is not an exhaustive list of applicable venues, we believe they are representative venues that are likely to include most communities and thereby reducing the risk with a snowball approach of missing relevant publications, as intended with the hybrid method. 

\begin{table}[]
\caption{Journals and conferences included in finding the Start set II. }
\label{tab:conf_journal}
\begin{tabular}{ll}
\toprule
Topic &  Forum \\ \midrule
 Software  & International Conference on Software Engineering \\
  Engineering & Asia-Pacific Software Engineering Conference \\
  & International Computer Software and Applications Conference \\
  & Symposium on Applied Computing \\
  & Transactions on Software Engineering \\
  & Information and Software Technology \\
  & Journal of Systems and Software \\
  Requirements  & International Requirements Engineering Conference \\
 Engineering & Requirements Engineering: Foundation for Software Quality\footnote{Actual text found in footnote text below.} \\
 & Requirements Engineering Journal\\
 Empirical Software  & Empirical Software Engineering Symposium\\
 Engineering & Empirical Software Engineering Journal \\
\bottomrule
\end{tabular}
\end{table}
\footnotetext{Before 2005, the REFSQ conference is not searchable in Scopus. As we are snowballing, we do not see this as a large threat to the validity. }

We used Scopus to search. The search was performed in September 2019. Table~\ref{tab:searchstring} outlines the components of the search string. The title and abstract were included in the search and only papers that include both the keyword for quality requirements as well as the keywords for empirical research. 

\begin{table}[]
\caption{The components of the search string used in Scopus. Scopus handles stemming, variation such as "non-functional" and "non functional" etc. Hence, possible variations are handled. }
\label{tab:searchstring}
\begin{tabular}{ll}
\toprule
Topic                   & Keyword  \\ \midrule
Quality requirements    & ("Quality requirements" OR \\ 
                        & "Non-functional requirements" OR \\ 
                        & "Extra functional properties") \\ \midrule
                        & AND \\ \midrule
Empirical studies       & ("Empirical" OR \\ 
                        & "Survey" OR \\ 
                        & "Case study" OR \\
                        & "Experiment" OR \\
                        & "Interviews" )\\
\bottomrule
\end{tabular}
\end{table}

\subsubsection{III. Snowballing}
The last step in our hybrid systematic review~\cite{mourao2017investigating} is the snowballing of the start set papers. We snowballed on the extended start set -- the combination of Start set I and Start set II -- to get to our final set of papers (cf. Figure~\ref{fig:process}). In a snowball approach, both references in the paper (backward references) and papers referring to the paper (forward references) were screened~\cite{wohlin2014guidelines}. We used Google Scholar to find forward references. 

\subsection{Planning}\label{sec:Planning}

We arrived at the following inclusion criteria, in a discussion among the researchers and based on related work:
\begin{enumerate}
    \item The paper should be on quality requirements or have quality requirements as a central result.
    \item There should be empirical results with a well-defined method and validity section, not just an example or anecdotal experience.
    \item Papers should be written in English. 
    \item The papers should be peer-reviewed. 
    \item The papers should be primary studies. 
    \item Conference or journal should be listed in reference ranking such as SJR. 
\end{enumerate}

Similarly, we defined our exclusion criteria as:
\begin{enumerate}
    \item Literature reviews, meta-studies, etc., -- secondary studies -- are excluded.
    \item If a conference paper is extended into a journal version, we only include the journal version.  
    \item We include papers only once, i.e., duplicates are removed throughout the process. 
    \item Papers focusing on only specific aspect(s) (Security, sustainability, etc.) are excluded.  
\end{enumerate}

All researchers were involved in the screening and classification process, even though the primary researcher performed the bulk of the work. The screening and classification were performed as follows: 

\begin{enumerate}
    \item Screen based on title and/or abstract. 
    \item We performed a full read when at least one researcher wanted to include the paper from the screening step.  
    \item Papers were classified according to the review protocol, see Section~\ref{sec:classification}. This was performed by the primary researcher and validated by another researcher. 
\end{enumerate}

To ensure reliability in the inclusion of papers and coding, the process was performed iteratively according to the sets. 
    \begin{enumerate}
        \item[I.] For Start set I, all references from the systematic literature review~\cite{BerntssonSvensson2010261} and systematic mapping study~\cite{ambreen2018empirical} were screened by two or three researchers. We only used the title in the screening step for Start set I. Full read and classification were performed by two or three researchers. 
        \item[II.] For Start set II, the screening was primarily performed by the primary researcher but with frequent alignment with at least one more researcher to ensure consistent screening -- both on title and abstract. Similarly for the full read and classification of the papers. Specifically, we paid extra attention to which papers to exclude to ensure we did not exclude relevant papers. 
        \item[III.] The primary researcher performed the snowballing. We screened on title only for backward and forward snowballing. We included borderline cases, to ensure we did not miss any relevant references. 
\end{enumerate}{}

The full read and classification were primarily performed by the primary researcher for Start set II and the Snowballing set. A sample of the papers was read by another researcher to improve validity in addition to those cases already reviewed by more than one researcher. 

The combined number of primary studies from the systematic literature review~\cite{BerntssonSvensson2010261} and systematic mapping study~\cite{ambreen2018empirical} are 288 in Start set I. However, there is an overlap between the two studies and there are some duplicates in the Ambreen et al. paper~\cite{ambreen2018empirical}. In the end, we had 274 unique papers in Start set I. After the screening, 41 papers remained. After the full read, additional papers were excluded resulting in 14 papers in the Start set I.  

Our search in Start Set II resulted in 190 papers. 173 papers remained after removing duplicate papers and papers already included in Start set I. After the screening and full read, the final Start set II was 23. Hence, the extended start set (combining Start set I and Start set II) together resulted in the screening of 447 papers and the inclusion of 37 papers. 

The snowball process was repeated until no new papers are found. We iterated 2 times -- denoted I1 and I2 in Figure~\ref{fig:process}. In iteration 1, we reviewed 77 papers and included 43. In iteration 2, we reviewed 6 papers and included 4. This resulted in a total of 84 papers included and 530 papers reviewed. 

\subsection{Classification and Review protocol}\label{sec:classification}
We developed the review protocol based on the systematic literature review~\cite{BerntssonSvensson2010261} and systematic mapping study~\cite{ambreen2018empirical}, the methodology papers~\cite{Kitchenham07guidelinesfor,wohlin2014guidelines}, and our research questions. The main items in our protocol are: 
\begin{itemize}
    \item Type of empirical study according to Wieringa et al.~\cite{wieringa2006requirements}. As we are focusing on empirical studies, we use the evaluation type -- investigations of quality requirements practices in a real setting --, validation type -- investigations of solution proposals before they are implemented in a real setting --, or experience type -- studies where the researchers are taking a more part in the study, not just observing. 
    \item Method used in the papers. We found the following primary methods used: Experiment, test, case study, survey, and action research.
    \item Analysis of rigor according to Ivarsson and Gorschek~\cite{ivarsson2011method}. 
    \item Thematic analysis of the papers -- in an initial analysis based on the author keyword and in later iterations further refined and grouped during the analysis process. 
\end{itemize}

We used a spreadsheet for documentation of the classification and review notes. The classification scheme evolved iteratively (see Section~\ref{sec:Planning}) as we included more papers. The initial themes were documented in the review process. In the analysis phase, the initial themes were used for an initial grouping of the papers. The themes were aligned and grouped in the analysis process of the papers, which included a number of meetings and iterative reviews of the results. The final themes which we used for the papers is the results of the iterative analysis process, primarily performed by the first and third researcher. 

\subsection{Validity}\label{sec:validity}
All cases where there were uncertainties whether to include a paper -- both in the screening step and the full read step -- or on the classification were reviewed by at least two researchers. Furthermore, to ensure consistent use of the inclusion and exclusion criterion as well as the classification we also sampled and reviewed papers that had only been screened or reviewed by only one researcher. 

We used Scopus for Start set II. We confirmed that all journals and conferences selected from Start set I were found in Scopus. However, REFSQ was only indexed from 2005 and onward. However, we do not see this as a problem as we are snowballing and the papers that are missing from the Scopus search due to this, should appear in the results through the snowballing process.

We used Google Scholar in the snowballing. This is recommended~\cite{wohlin2014guidelines} and usually gives the most complete results.

A hybrid search strategy can be sensitive to starting conditions, as pointed out by Mourão et al.~\cite{mourao2017investigating}. However, their results indicate that the strategy can produce similar results as a standard systematic literature review. We carefully selected the systematic literature review and the systematic mapping study as Start set I and extended it with a keyword-based search for selected forums in Start set II. Hence, we believe the extended start set on which we snowballed is likely to be sufficient to ensure a good result when complemented with the snowball approach.


\section{Analysis and Results}\label{sec:results}
\begin{figure}
\centering
\includegraphics[width=0.99\textwidth]{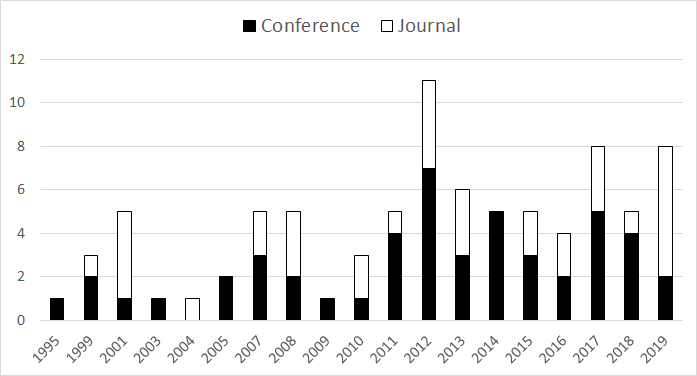}
 \caption{An overview of papers included -- publication type and year of publication. }
  \label{fig:type-year}
\end{figure}

The screening and reading of the papers in Start set I was performed in August and September 2019. The keyword-based search for Start set II was performed in September 2019. The snowballing was subsequently performed in October and November. In total, 530 papers are screened, of which 194 papers are read in full. This resulted in including 84 papers, from 1995 to 2019 -- see Figure~\ref{fig:type-year}. 

\subsection{RQ1 Which empirical methods are used to study quality requirements?}\label{sec:overall_results}

The type of studies performed are found in Table~\ref{tab:method_overview} -- categorized according to Wieringa et al.~\cite{wieringa2006requirements}. We differentiate between two types of validations: experiments involving human subjects and tests of algorithms on a data set. For the latter, the authors either report experiment or case study, whereas we call them test. The evaluations we found are either performed as case studies or surveys. Lastly, we found three papers that used action research -- categorized as experience in Table\ref{tab:method_overview}. It should be noted that the authors of the action research papers did not themselves explicitly say they performed an action research study. However, when we classified the papers, it is quite clear that, according to Wieringa et al.~\cite{wieringa2006requirements}, they are in the action research category.

\begin{table}[]
\caption{Type of empirical study and context. }
\label{tab:method_overview}
\begin{tabular}{l|cc|cc|c|c}
\toprule
Type        & \multicolumn{2}{c}{Validation} & \multicolumn{2}{c}{Evaluation}        & \multicolumn{1}{c}{Experience}      &  \\ 
Method      & Experiment        & Test       & Case study & Survey & Action research & Total      \\ \midrule
Academic    & 4                 & 13         & 5          &        & 1               & 23    \\
Industry    &                   & 4          & 35         & 14     & 2               & 55    \\
Mixed       &                   & 2          & 2          & 1      &                 & 5     \\
Open source &                   &            & 1          &        &                 & 1     \\ \midrule
Total       & 4                 & 19         & 43         & 15     & 3               & 84   \\ \bottomrule
\end{tabular}
\end{table}

\begin{figure}
\centering
\includegraphics[width=0.99\textwidth]{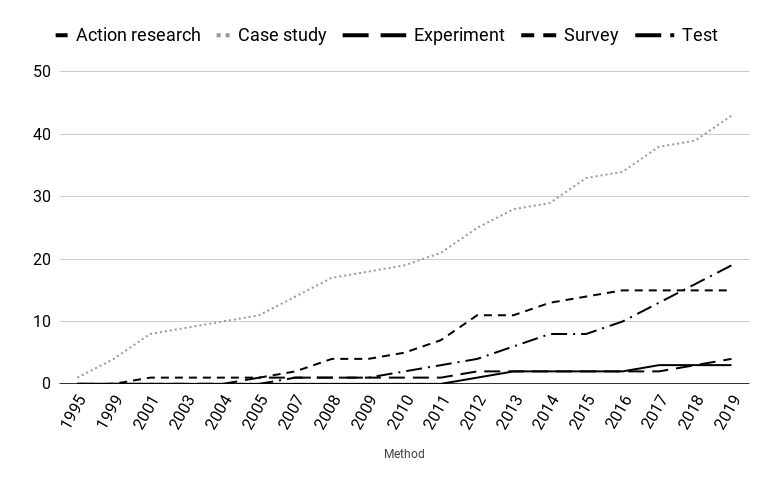}
 \caption{An overview of papers included -- method and accumulated number of publications per year. }
  \label{fig:method-year}
\end{figure}

Case studies in an industry setting are the most common (35 of 84), followed by surveys in industry (14 of 84) and test in academic settings (13 of 84). This indicates that research on quality requirements is applied and evidence is primarily of individual case studies rather than through validation in laboratory settings. Case studies seem to have similar popularity over time, see Figure~\ref{fig:method-year}. We speculate that since requirements engineering in general as well as quality requirements in particular is a human-intensive activity, there are not so many clear cause-effect relationships to study in a rigorous experiment. Rather, it is more important to study practices in realistic settings. However, there are only three longitudinal studies. 

Tests are, in contrast to the case studies, primarily performed in an academic setting, which is not necessarily representative in terms of scale and artifacts. The papers are published from 2010 to 2019 -- one exception, published in 2007, see Figure~\ref{fig:method-year}. One explanation might be the developments in computing driving the trend to use large data sets.

We found only one study on open source, see Table~\ref{tab:method_overview}, which is also longitudinal. We speculate that requirements engineering is sometimes seen as a business activity where someone other than the developers decides what should be implemented. In open-source projects, there is often a delegated culture where there is no clear product manager or similar deciding what to do, albeit there can be influential individuals such as the originator or core team member. We believe this entails that quality requirements engineering is different in open source projects than when managed within an organization. It would be interesting to see if this hypothesis holds for requirements engineering in general and not just quality requirements. We believe, however, that by studying forums, issue management, and reviews that open source projects are an untapped resource for quality requirements research. 

We classify the studies according to rigor, as proposed by Ivarsson and Gorschek~\cite{ivarsson2011method}. We assess the design and validity rigor. Table~\ref{tab:overview_rigour} presents our evaluation of design and validity rigor in the papers. An inclusion criterion is that there should be an empirical study, not just examples or anecdotes. Hence, it is not surprising that overall studies score well in our rigor assessment.

\begin{table}[]
\caption{Overview of rigor in the studies, based on Ivarsson and Gorschek~\cite{ivarsson2011method}. We evaluate the design description -- whether it would be possible to replicate the study -- and validity description -- whether the relevant validity threats are well described. A score of "1" means the design or validation is well-described in sufficient detail. A score of "0.5" means the design or validation is briefly described but lacking relevant details. A score of "0" means there is no description or it is described in such a way that it is not possible to understand the details. }
\label{tab:overview_rigour}
\begin{tabular}{cc|ccc|c}
\toprule
Design score  & Validation score & Validation & Evaluation & Experience      &  Total     \\  \midrule
1        & 1         & 11          & 25           & 1               & 37    \\
          & 0.5   &                  1          & 5                             & 1               & 7     \\
          & 0                            & 5          & 5           &                 & 10    \\ \midrule
0.5 & 1                           & 1                           & 1           &                 & 2     \\
          & 0.5     &                              & 5           &                 & 5     \\
          & 0          & 1                            & 10           &                 & 11    \\ \midrule
0        & 0                            & 4          & 7                             & 1               & 12    \\ \midrule
Total     &             & 23         & 58          & 3               & 84    \\ \bottomrule
\end{tabular}
\end{table}

High rigor is important for validations studies -- to allow for replications -- which is also the case for 11 out of 23 studies. The number increases to 13 if we include papers with a rigor score 0.5 for both design and validity and to 19 if we focus solely on the design rigour. Interestingly, we found no replication studies. Furthermore, the number of studies on a single (similar) approach or solution is in general low. We speculate that the originators of a solution have an interest in performing empirical studies on their solution. However, it seems unusual that practitioners or empiricists with no connection to the original solution or approach try to apply it. Furthermore, we also speculate that academic quality requirements research is not addressing critical topics for industry as there seems not to be an interest in applying and learning more about them. This implies that the research on quality requirements might need to better understand what are the real issues facing software developing organizations in terms of quality requirements. 

The validity part of rigor is also important for evaluations and experience papers. Strict replications are typically not possible. However, understanding the contextual factors and validity are key in interpreting the results and assessing their applicability in other cases and contexts. 22 of the 58 evaluation and 1 of the 3 experience papers do not have a well-described validity section (rigor score 0), and 10 evaluation and 1 experience paper have a low score (rigor score 0.5). Hence, we conclude that the overall strength of evidence is weak. 

\subsubsection{Validations}\label{sec:results_validation}
Experiments are, in general, the most rigorous type of empirical study with the most control. However, it is difficult to scale to a realistic scenario. We found four experiments validating quality requirements with human subjects, see Table~\ref{tab:Experiments}.

\begin{table}
    \caption{Overview of the experiment studies.}
    \label{tab:Experiments}
    \begin{tabular}{@{}lllll@{}}
        \toprule
        Reference       & Year              & Theme     & Subjects & Context  \\ \midrule
        \citeSR{SLR268} & 2005              & ISO9126    & 158      & Students \\
        \citeSR{SLR619} & 2012              & CSRML (i*) & 84       & Students \\
        \citeSR{SLR575} & 2018              & Templates  & 107      & Students \\ 
        \citeSR{SLR571} & 2019              & i*         & 32       & Students \\ \bottomrule
    \end{tabular}
\end{table}

We note that all experiments are performed with students -- at varying academic levels. This might very well be appropriate for experiments~\cite{host2000EMSEstudentsassubjects}. We notice that there are only four experiments, which might be justified by: 1) Experiments as a method is not well accepted nor understood in the community. 2) Scale and context are key factors for applied fields such as requirements engineering, making it more challenging to design relevant experiments. 

Several empirical studies study methods or tools by applying them to a data set or document set. We categorize those as tests, see Table~\ref{tab:tests-overview}. We found three themes for tests. 
\begin{enumerate}
    \item Automatic analysis - the aim is to evaluate an algorithm or statistical method to automatically analyze a text, usually a requirements document.
    \item Tool - the aim is to evaluate a tool specifically.
    \item Runtime analysis - evaluating the degree of satisfaction of quality requirements during runtime.  
\end{enumerate}

Tests are also fairly rigorous in that it is possible to control the study parameters well. It can also be possible to perform realistic studies, representative of real scenarios. The challenge is often to attain data that is representative. The dataset are described in Table~\ref{tab:test-dataset}.  

\begin{table}[]
\caption{Overview of the test studies. The dataset are described in Table~\ref{tab:test-dataset}.}
\label{tab:tests-overview}
\begin{tabular}{llll}
\toprule
Ref.            & Year & Dataset                                    & Theme             \\ \midrule
\citeSR{SLR137} & 2007 & DePaul07, Siemens IET                      & Automatic analysis \\
\citeSR{SLR227} & 2010 & DePaul07                                   & Automatic analysis \\
\citeSR{SLR242} & 2011 & DePaul07                                   & Automatic analysis \\
\citeSR{SLR669} & 2012 & DePaul07, EU procurement                   & Tool               \\
\citeSR{SLR246} & 2013 & DePaul07, CCHIT, iTrust, openEMR, DUA, RFP & Automatic analysis \\
\citeSR{SLR241} & 2013 & HWS, CRS (OSS projects)                    & Tool               \\
\citeSR{SLR228} & 2014 & CCHIT, WorldVista                          & Automatic analysis \\
\citeSR{SLR249} & 2014 & DePaul07                                   & Automatic analysis \\
\citeSR{SLR243} & 2016 & DePaul07, Concordia corpus                 & Automatic analysis \\
\citeSR{SLR657} & 2016 & Proprietary                                & Automatic analysis \\
\citeSR{SLR29}  & 2017 & DePaul07                                   & Automatic analysis \\
\citeSR{SLR248} & 2017 & Mobile app store reviews                   & Automatic analysis \\
\citeSR{SLR261} & 2017 & PTC, MIP, CCHIT                            & Automatic analysis \\
\citeSR{SLR277} & 2018 & DePaul07                                   & Automatic analysis \\
\citeSR{SLR665} & 2018 & DePaul07                                   & Automatic analysis \\
\citeSR{SLR6}   & 2018 & TAS, deletaIoT                             & Runtime analysis             \\
\citeSR{SLR245} & 2019 & Mobile app store reviews                   & Automatic analysis \\
\citeSR{SLR278} & 2019 & DePaul07, Predictor models                 & Automatic analysis \\
\citeSR{SLR235} & 2019 & DePaul07, CCHIT                            & Automatic analysis \\ \bottomrule
\end{tabular}
\end{table}

The most commonly used data set it the DePaul07 data set~\cite{dePaul07}. It consists of 15 annotated specifications from student projects at DePaul University from 2007. This data set consists of requirements specification -- annotated to functional and quality requirements as well as the type of quality requirement -- from student projects.  

\begin{table}[]
\caption{Data sets. In most cases when they are openly accessible, they can, with some effort, be found online. Question marks indicate that we were unable to identify the necessary information, neither in the referred paper nor from the resources indicated in the papers. }
\label{tab:test-dataset}
\begin{tabular}{@{}llll@{}}
\toprule
Data set                 & Context  & Access & Content                                        \\ \midrule
DePaul07                 & Academic & Open         & Annotated requirements \cite{dePaul07} \\
Siemens IET              & Industry & Closed       & Requirements document \citeSR{SLR137}                                         \\
EU procurement           & Public   & Open         & Requirements document \citeSR{SLR669}                                     \\
CCHIT                    & Public   & Open         & Requirements document \citeSR{SLR246}              \\
iTrust                   & ?        & ?            & Requirements documents \citeSR{SLR246}                                         \\
OpenEMR                  & Public   & Open         & Manuals~\citeSR{SLR246}                                         \\
DUA                      & Public   & ?            & Data Use Agreements~\citeSR{SLR246}                                         \\
RFP                      & Public   & ?            & Request for Proposals~\citeSR{SLR246}                                         \\
HWS                      & ?        & ?            & ?~\citeSR{SLR241}                                \\
CRS                      & Example  & Open         & Use case documents~\citeSR{SLR241}                                \\
WorldVistas              & Public   & Open         & Requirements document~\citeSR{SLR228}                                               \\
Concordia RE corpus      & Mixed (A+P) & Open & Requirements documents~\citeSR{SLR243} \\
PTC                      & Industry & Closed       & Project documents~\citeSR{SLR261}                                    \\
MIP                      & Academic & Closed       & Project documents~\citeSR{SLR261}                                    \\
TAS                      & Example  & Open         & Project documents~\citeSR{SLR6}                               \\ 
deltaIoT                 & Example  & Open         & Project documents~\citeSR{SLR6}                               \\ 
Mobile app store reviews & Industry & Open         & App reviews                               \\ 
Predictor models         & ?        & ?            & ?                                \\ \bottomrule
\end{tabular}
\end{table}

There are few examples where data from commercial projects have been used. The data do not seem to be available for use by other researchers. There are examples where data from public organizations -- such as government agencies -- are available and used, e.g. the EU procurement specification, see Table~\ref{tab:test-dataset}. 

The most common type of data is a traditional requirements document, written in structured text. There are also a couple of instances where use case documents are used. For non-requirements specific artifacts, manuals, data use agreements, request for proposals (RFPs), and app reviews are used. From the papers in this systematic literature view, artifacts such as backlogs, feature lists, roadmaps, pull requests, or test documents do not seem to have been included. 

\subsubsection{Evaluations}\label{sec:results_evaluation}
\begin{figure}
\centering
\includegraphics[width=0.60\textwidth]{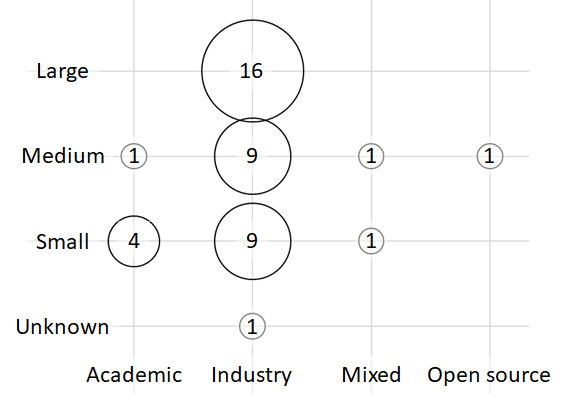}
 \caption{Overview of the case studies on quality requirements -- 43 papers of the 84 included in this literature review. Scale refers to the context of the case study -- small:  sampling parts of the context (e.g. one part of a company) or is overall a smaller context (e.g. example system), medium: sampling a significant part of the context or a larger example system, large: sampling all significant parts of a context of an actual system (not example or made up). The context is also classified according to where the case studies are executed. Academic means primarily by students (at some academic level). Mixed means the case studies are executed in both an academic and industry context. For details, please see Table~\ref{tab:casestudies_all}.}
  \label{fig:casestudies_overview}
\end{figure}

It is usually not possible to have the same rigorous control of all study parameters in case studies~\cite{easterbrook2008selecting}. However, it is often easier to have realistic scenarios, more relevant for a practical setting. We found both case studies performed in an industry context with practitioners as well as in an academic context with primarily students at different academic levels. We found 43 papers presenting case study reports on quality requirements, see Figure~\ref{fig:casestudies_overview} (all details can be found in Table~\ref{tab:casestudies_all}). We separate case studies that explicitly evaluate a specific tool, method, technique, framework, etc., and exploratory case studies aiming to understand a specific context rather than evaluating something specific. 

Case studies sometimes study a specific object, e.g., a tool or method, see Table~\ref{tab:casestudies_all}. We found 25 case studies explicitly studying a particular object. Two objects are evaluated more than once, otherwise just one case study per object. We found no longitudinal cases; hence, the case studies are executed at one point in time and not followed up at a later time. The QUPER method is studied in several case studies in several different contexts (see Table~\ref{tab:casestudies_all}). There are several case studies for the NFR method, however, it seems the context is similar or the same in most of the cases (row 2). 

We found 18 exploratory case studies on quality requirements where a specific object wasn't the focus, see Table~\ref{tab:casestudies_all}. Rather, the goal is to understand a particular theme of quality requirements. 8 case studies want to understand details of quality requirements, e.g., the prevalence of a specific type of quality requirement or what happens in the lifecycle of a project. 5 case studies have studied the process around quality requirements, two studies on sources of quality requirements (in particular app reviews), two studies in particular on developers' view on quality requirements (specifically using StackOverflow), lastly one study on metric related to quality requirements. We found two longitudinal case studies. 

The goal of a survey is to understand a broader context without performing any intervention~\cite{easterbrook2008selecting}. Surveys can be used either very early in the research process before there is a theory to find interesting hypotheses or late in the process to understand the prevalence of a theme from a theory in a certain population. We found 15 surveys, see Table~\ref{tab:surveys_overview}; 5 interviews, 9 questionnaires, one both. The goals of the surveys are a mix of understanding practices around the engineering of quality requirements and understanding actual quality requirements as such.

\begin{table}[]
\caption{Overview of the surveys found. The surveys in our study either use interviews or questionnaires -- or both -- to collect data. Size refers to how many participants -- respondents to questionnaire or interviewees in interviews.  }
\label{tab:surveys_overview}
\begin{tabular}{@{}llllll@{}}
\toprule
Ref.            & Year & Type          & Size      & Perspective        & Theme                              \\ \midrule
\citeSR{SLR219} & 2008 & Interviews    & 5         & Development        & Importance, Practices              \\
\citeSR{SLR89}  & 2011 & Interviews    & 11        & Development        & Prioritization                     \\
\citeSR{SLR76}  & 2012 & Interviews    & 13        & Architects         & Importance, Practices              \\
\citeSR{SLR81}  & 2012 & Interviews    & 22        & Development        & Importance, impact                 \\
\citeSR{SLR57}  & 2014 & Interviews    & 16        & Architects         & Outsourcing, Alignment \\
\citeSR{SLR264} & 2014 & Both          & 48\footnote{14 interviews and 34 questionnaire respondents.} & Development        & Importance, Practices              \\
\citeSR{SLR290} & 2001 & Questionnaire & 34        & Development        & Alignment, Importance  \\
\citeSR{SLR653} & 2007 & Questionnaire & 75        & Users              & User satisfaction, Importance      \\
\citeSR{SLR120} & 2008 & Questionnaire & 133       & Procurement        & Importance                         \\
\citeSR{SLR539} & 2010 & Questionnaire & 318       & Development, Users        & Importance                         \\
\citeSR{SLR258} & 2011 & Questionnaire & 35        & Development        & Importance                         \\
\citeSR{SLR203} & 2012 & Questionnaire & 39        & Development        & Importance            \\
\citeSR{SLR664} & 2012 & Questionnaire & 29        & Development        & Quality model         \\
\citeSR{SLR273} & 2015 & Questionnaire & 53        & Development        & Importance, Practices \\
\citeSR{SLR259} & 2016 & Questionnaire & 56        & Development        & Importance            \\ \bottomrule
\end{tabular}
\end{table}
\footnotetext{14 interviews and 34 questionnaire respondents.}

Overall, the surveys we found are small in terms of the sample of the population. In most cases, they do not report from which population they sample. The most common theme is the importance of quality requirements and specific sub characteristics -- typically according to ISO9126~\cite{ISO9126} or ISO25010~\cite{ISO25010}. However, we cannot draw any conclusions as sampling is not systematic and the population unclear. We believe it is not realistic to systematically sample any population and achieve a statistically significant result on how important quality requirements are nor which sub characteristics are more or less important. We speculate that, besides the sampling challenge, the variance among organizations and point in time will likely be large, making the practical implications of such studies of questionable value. 

\subsubsection{Experience}\label{sec:results_experience}
In action research, the researchers are more active and part of the work than e.g. in a case study~\cite{easterbrook2008selecting}. Whereas a case study does not necessarily evaluate a specific object, action research typically reports some kind of intervention where something is changed or performed. We found 3 papers we classify as action research types of experience papers~\cite{wieringa2006requirements}, see Table~\ref{tab:actionresearch_overview}. 

Two of the studies are performed at one point in time~\citeSR{SLR239,SLR253}. One study is longitudinal, describing the changes to the processes and practices around quality requirements over several years~\citeSR{SLR222}. Interestingly, all three studies directly refer to ISO9126~\cite{ISO9126} or ISO25010~\cite{ISO25010}. 

\begin{table}[]
\caption{Action research}
\label{tab:actionresearch_overview}
\begin{tabular}{@{}lllp{7cm}@{}}
\toprule
Reference     & Year & Context             & Theme \\ \midrule 
\citeSR{SLR239} & 2012 & Industry (Siemens)  & Quality models and ISO9126.                     \\
\citeSR{SLR253} & 2013 & Public (University) & Using ElicitO to help intranet development at the university. \\
\citeSR{SLR222} & 2017 & Public (gov ag)     & Working several years to improve processes, using ISO9126.    \\ \bottomrule 
\end{tabular}
\end{table}

\subsection{RQ2 What are the problems and challenges for quality requirements identified by empirical studies?}
We have grouped the studies on quality requirements themes thematically to analyze the problems and challenges that are identified in the empirical studies. The groups are developed iteratively among the researchers, initially from the author keywords in the included papers and then iteratively refined. 

\subsubsection{Quality requirements and other requirements}
There is an academic debate on what quality requirements are and what they should be called~\cite{glinz2007non}. We found a study indicating that quality requirements -- sometimes called non-functional requirements -- are functional or behavioural~\citeSR{SLR36}. This is in line with other studies that report that a mix of requirements type is common~\citeSR{SLR215}. Two studies find that architects address quality requirements the same way as other requirements~\citeSR{SLR201,SLR203}, also confirmed in other surveys~\citeSR{SLR231}. However, there are also research studies indicating a varying prevalence and explicitness than other requirements~\citeSR{SLR36,SLR274,SLR215,SLR217,SLR224}. We interpret the current state of evidence to be unclear on the handling of quality requirements. We speculate that the answers to opinion surveys might be biased towards the expected "right" answer -- as expected by the researcher -- rather than the actual viewpoint of the respondent. 

\subsubsection{Importance and prevalence of quality requirements}
Many papers present results related to the importance and prevalence of quality requirements -- or sub characteristics of quality requirements. Four papers present results from artifact analysis~\citeSR{SLR215,SLR274,SLR667,SLR217}.  
We found eight personal opinion survey papers ~\citeSR{SLR81,SLR120,SLR76,SLR57,SLR258,SLR259,SLR273,SLR264}.  
Similarly, a list of quality requirements types is developed through a survey for service-oriented applications~\citeSR{SLR664}.   
Furthermore, we found three papers analyzing app store reviews~\citeSR{SLR237,SLR245,SLR248} and two papers developer's discussions on StackOverflow~\citeSR{SLR182,SLR185} and one paper studying 8 open source projects communication~\citeSR{SLR236}.  
The individual papers do not present statistical tests or variance measures. Furthermore, we found no papers elaborating on a rationale for why the distribution of sub characteristics of quality requirements are more or less prevalent or seems as important by the subjects. We hypothesize that the importance of different quality requirements types varies over time, domain, and with personal opinion. This implies that there is no general answer to the importance of different quality requirements types. Rather, we believe it is important to adapt the quality requirements activities -- such as planning and prioritization -- to the specific context rather than to use predefined lists. 

\subsubsection{Specification of quality requirements}
We found three case study papers reporting on artifact analysis of realistic requirements documents~\citeSR{SLR36,SLR215,SLR217}. The practice seems to vary in how quality requirements are written; quantification, style, etc. One paper reporting on a scope decision database analysis~\citeSR{SLR274}. The prevalence of quality requirements features is low and varies over time. Two interview surveys, furthermore, find quantification varies for the different cases as well as for the different quality requirements types~\citeSR{SLR81,SLR201}. Four surveys indicate that quality requirements are often poorly documented and without template~\citeSR{SLR76,SLR219,SLR264,SLR263}. Overall, the studies mostly report the usage of informal specification techniques (structured text) rather than specific modelling notations. 

\subsubsection{Roles perspective}
Different roles -- for example, project manager, architect, product manager -- view and work with quality requirements differently. Two interview surveys report that architects are often involved in the elicitation and definition of quality requirements~\citeSR{SLR76,SLR201}. Furthermore, the clients or customers -- in a bespoken context -- are not explicit nor active in the elicitation and definition of quality requirements~\citeSR{SLR201}. We found six papers collecting opinion data on the priority of quality requirements types from a role perspective~\citeSR{SLR258,SLR81,SLR290,SLR259,SLR76,SLR539}. We did not find any particular trend nor general view for different roles, except that when asked subjects tend to answer that quality requirements as a topic is important and explicitly handled -- albeit that there are improvement potentials. Hence, it seems to us that, again, there might not be a general answer to the importance of different quality requirements types. 

One study found that architects -- despite being one source of quality requirements -- are not involved in the scoping~\citeSR{SLR57}. Another study found that relying on external stakeholders might lead to long lead-times and incomplete quality requirements~\citeSR{SLR274}. We found one study on quality requirements engineering in an agile context. They report that communication and unstated assumptions are major challenges for quality requirements~\citeSR{SLR279}. Even though opinion surveys indicate that subjects -- independent of roles -- claim to prioritize and explicitly work with quality requirements, there are indications that implicit quality requirements engineering is common and this leads to misalignment. 

We find evidence of how different roles perceive and handle quality requirements to be insufficient to draw any particular conclusions. 

\subsubsection{Lifecycle perspective}
We found two papers presenting results related to changes over time for the prevalence of different quality requirements types. Ernst and Mylopoulos study 8 open source project~\citeSR{SLR236} and Olsson et al. study scope decisions from a company~\citeSR{SLR274}. Ernst and Mylopoulos did not find any specific pattern across the 8 open source project in terms of prioritization or scoping of quality requirements. Olsson et al. conclude that there was an increase in the number of quality-oriented features and the acceptance of quality requirements in the scope decision process later in the product lifecycle compared to early in the product lifecycle. We found one student experiment on the stability of prioritization within the release of a smaller project~\citeSR{SLR42}. They conclude that interoperability and reliability are more stable in terms of project priority whereas usability and security changed priority more in the release cycle. Lastly, we found a paper presenting a study on the presence of ``Not a Problem" issue reports in the defect flow compared to how precise quality requirements are written~\citeSR{SLR340}. The main result is that the more precise quality requirements are written, the lower the amount of ``Not a Problem" issue reports. 

The number of studies is small, which makes it difficult to draw any conclusions. However, we speculate that what happens over time is also likely to vary and be context-specific. We hypothesize that there might be general patterns that, for example, products early in the lifecycle tend to overlook quality requirements whereas products later in the lifecycle tend to focus more on quality requirements. Furthermore, it might also be differences in the handling of quality requirements depending on how close to release the project is. We see these topics as relevant to study in more detail. Longitudinal studies, involving different artifacts and sources information, e.g. issue report systems, can be an interesting way forward. 

\subsubsection{Prioritization}
We found two case studies on quality requirements prioritization~\citeSR{SLR89,SLR201}. Berntsson Svensson et al. conducted an interview study with product managers and project leaders~\citeSR{SLR89}. They found that ad-hoc prioritization and priority grouping of quality requirements are the most common. Furthermore, they found that project leaders are more systematic (55\% prioritize ad-hoc) compared to the product managers (73\% prioritize ad-hoc). Daneva et al. found in their interview study that architects are commonly involved in prioritization of quality requirements~\citeSR{SLR201}. They identified ad-hoc and priority grouping as the most common approach to prioritization. Daneva et al., furthermore, found that 7 out of the 20 architects they interviewed considered themselves the role that sets the priority for quality requirements. 

In summary, we find there is overall a lack of understanding of quality requirements prioritization. The studies indicate the involvement of different roles, which we believe warrants further research. Furthermore, the lack of systematic prioritization seems to be in line with requirements in general and not just for quality requirements. 

\subsubsection{Sources of quality requirements}
There can be several sources of requirements, both roles as well as artifacts. As reported before, architects are sometimes involved in the elicitation and definition of quality requirements~\citeSR{SLR76,SLR201}. Three studies have identified user reviews on mobile app markets as a potential source of quality requirements~\citeSR{SLR237,SLR245,SLR275}. One study found that users are not sufficiently involved in the elicitation~\citeSR{SLR224}. However, we did not find studies on, for example, usage data or customer services data as a means to elicit and analyze quality requirements.

\subsection{RQ3 Which quality requirements solution proposals have been empirically validated?}

Several techniques, tools, etc., have been proposed to address problem and challenges with quality requirements. The results of the evaluations or validation of different quality requirements solutions are grouped after similarity. The sections are ordered according to the number of studies. 

\subsubsection{Automatic analysis}
One research direction which has gained popularity is different forms of automatic analysis. The idea is that a tool can be developed to support human engineers in different aspects of quality requirements engineering. All studies we found reported positive results. 

We found a number of papers investigating automatic identification and classification of quality requirements from different types of sources~\citeSR{SLR29,SLR137,SLR227,SLR228,SLR235,SLR242,SLR243,SLR246,SLR249,SLR261,SLR277,SLR278,SLR657,SLR665}. The different papers test different algorithms and approaches on different data sets. The most commonly used data set is from a project course at DePaul University from 2007. That data set has annotated requirements (functional or quality requirements) as well as quality requirements types, see Table~\ref{tab:tests-overview} and Table~\ref{tab:test-dataset}. Overall, the studies are executed in an academic context (11 out of 14) and all at a small scale which might not be representative for realistic commercial cases. Furthermore, it is often assumed the presence of requirements documents, which might not be the case for agile contexts. 

We found two papers presenting studies of user reviews in app stores, e.g. Apple app store or Google play~\citeSR{SLR245,SLR248}, both rigorous. Similar to other work on automatic classification, the two studies evaluated different algorithms to identify and classify quality requirements in app reviews. 

We found one paper on early aspect mining~\citeSR{SLR241}. The study evaluated a tool to detect quality requirements aspects in a requirements document. Based on the detection, quality requirements are suggested to the requirements engineer. Another study evaluated a use case tool with explicit quality requirements for an agile context~\citeSR{SLR669}. Both studies imply feasibility but cost or amount of effort of using them in large-scale realistic cases are not studied. 

We found one study on runtime adaptations of quality requirements for self-managing systems~\citeSR{SLR6}. Rather than defining fixed values for trade-offs among quality requirements, quality requirements are defined as intervals that can be optimized in runtime, depending on the specific operational conditions. They test their approach on two example systems, which show better compliance when using their approach than not. 

We summarize that, while there are many tests and experiments, there are few studies of realistic scale and with realistic artifacts on automatic analysis in an quality requirements context. We also find that there is a lack of costs of running the automatic analysis, such as preparation of data, needs in terms of hardware and software, and knowledge needed by an analyst. We conclude that automatic analysis shows promise in an academic setting but has yet to be studied in a realistic scale case study or action research. 

\subsubsection{Goal modeling}
We found two experiments on different extensions of i*, validating usefulness and correctness of the extensions compared to the original i* approach~\citeSR{SLR619,SLR571}. The experiments are conducted in an academic setting. Both experiments conclude that the extensions are better. Based on these experiments, we cannot say anything in general about modeling of quality requirements and usefulness of i* in general. 

Researchers have performed several case studies~\citeSR{SLR170,SLR162,SLR223,SLR284}. The researchers and case context are similar and all present how the NFR method and goal modeling can work in different situations. One case study evaluated a process where business models and a quality requirements catalog are used to finally build a goal model for relevant quality requirements~\citeSR{SLR252}. We found one case study using goal modeling to support product family quality requirements using goal modeling~\citeSR{SLR341}. All of these case studies are of low rigor both in terms of design and validity. 
We have found one paper describing a test of generating goal graphs from textual requirements documents~\citeSR{SLR228}, and another paper testing a tool for goal modeling in an agile context~\citeSR{SLR669}. Both papers indicate feasibility, i.e., the techniques seem to work in their respective context. 

Overall, the evidence point to that goal modeling -- in various forms -- can be used and does add benefits in terms of visualization and systematic reasoning. However, we have not found any realistic scale case studies on quality requirements, nor any data on effort or impact on other parts of the development. We have not found any surveys on modeling techniques used for quality requirements. Hence, we have not found evidence of the use of goal modeling in industry specifically for quality requirements. We judge the collected evidence that goal modeling does have potential benefits but they have not been evaluated in realistic scale projects with a systematic evaluation of the whole quality requirements process. 

\subsubsection{Quality models and ISO9126 / ISO25010}

ISO9126~\cite{ISO9126} -- and the updated version in ISO25010~\cite{ISO25010} -- is used in many of the papers we found. Al-Kilidar et al. validate the usefulness of ISO9126 in an experiment with students~\citeSR{SLR268}. They conclude that the standard is difficult to interpret and too general to be useful. However, the experiment has a low rigor score -- 0.5 for design and 0 for validity -- and lacks relevant validity description. 

ADEG-NFR use ISO25010~\cite{ISO25010} as catalogue of quality requirements types~\citeSR{SLR233}. The IESE NFR method also uses the ISO standard as the basis for creating checklists and quality models~\citeSR{SLR285}. Sibisi and van Waveren proposes a similar approach as the IESE NFR method, using the ISO standard as a starting point in customizing checklists and quality model~\citeSR{SLR297}. Two papers present evaluations of two approaches combining ISO9126 quality models with goal modeling~\citeSR{SLR252,SLR253}. Another paper evaluated an approach where a checklist was derived from ISO9126 to guide the elicitation~\citeSR{SLR291}. Similarly, two papers evaluate workshop and brainstorming approaches to elicitation and analysis based on ISO9126, which they propose to complement with multi-stakeholder workshops~\citeSR{SLR230,SLR661}. Lastly, the Quamoco approach suggests connecting the abstract quality model in the ISO standard with concrete measurements~\citeSR{SLR662}. 

We found one paper defining a catalog of quality requirements for service-oriented applications~\citeSR{SLR664}. They proposed an initial list of quality requirements which was evaluated with practitioners using a questionnaire-based approach. Mohagheghi and Aparicio conducted a three-year-long project at the Norwegian Labour and Welfare Administration~\citeSR{SLR222}. The aim was to improve the quality requirements. Lochmann et al. conducted a study at Siemens, Germany~\citeSR{SLR239}. The business unit in question develops a traffic control system. They introduced a quality model approach to the requirements process for quality requirements. The approach is based on ISO9126~\cite{ISO9126}. 

All of the approaches suggest incorporating the quality requirements specific parts into the overall requirements process as well as tailoring to the needs of the specific organization. The different approaches seem to be recognized as useful in realistic settings, leading to a more complete understanding of the quality requirements scope with a reasonable effort. It further seems as if the tailoring part is important to gain relevance and acceptance from the development organizations -- especially when considering the experiment from Al-Kilidar et al.~\citeSR{SLR268}. 

\subsubsection{Prioritization and release planning}
We found one method -- QUPER -- focused explicitly on prioritization and release planning~\citeSR{SLR296}. The researchers behind the method have performed several case studies to evaluate it~\citeSR{SLR281,SLR91,SLR197,SLR79}. We also found a prototype tool evaluation~\citeSR{SLR91}. This is the single most evaluated approach. QUPER is the only approach we found which is explicitly focused on prioritization and release planning. 

We found one paper proposing and evaluating an approach to handle interaction and potentially conflicting priorities among quality requirements~\citeSR{SLR272}. This is similar to QUARCC, which is tested in a tool evaluation~\citeSR{SLR287}. 

We summarize that QUPER has been evaluated both with academics and with practitioners in realistic settings. However, the long-term impact of using QUPER seems not to have been studied. However, other than QUPER, we conclude that there is no strong evidence for other solutions for prioritization and release planning of quality requirements. 

\subsubsection{Metrics and quality models}
We found two papers evaluating the connection between key metrics to measure quality requirements types and user satisfaction~\citeSR{SLR293,SLR653}. The results imply that quality requirements metrics -- measuring the presence of quality requirements types according to ISO9126~\cite{ISO9126} in the specifications -- is correlated with user satisfaction. Hence, even though this is not a longitudinal study, there are implications that good quality requirements engineering practices might increase user satisfaction. Both studies are personal opinion surveys, which makes it difficult to evaluate causality and root-cause. Furthermore, they measured at one point in time. 

We found one paper proposing and evaluating an approach to create metrics to evaluate the responses to a request for proposal (RFP)~\citeSR{SLR547}. The metrics are based on recommendations from authorities and focused on process metrics rather than product metrics. They report in their case study that they could identify specifications with deficiencies in quality requirements. 

We summarize that there is not a lot of evidence on the usage of metrics in connection to quality requirements. We find that the studies have identified interesting hypotheses that can be evaluated both in an academic setting through experiments or case studies as well as in real settings through case studies or action research. 

\subsubsection{Knowledge management}
We found three papers on different aspects of knowledge management to address quality requirements engineering. Balushi et al. report on a study at the University of Manchester~\citeSR{SLR253}. They applied the ElicitO framework on a project to enhance the university website. The ontology in ElicitO implements ISO9126~\cite{ISO9126}. The MERliNN framework suggests procedures to identify and manage knowledge flows in the elicitation and analysis process~\citeSR{SLR267}. One paper evaluating a tool for QUARCC and S-Cost -- knowledge-based tools for handling inter-relationships among quality requirements and stakeholders~\citeSR{SLR287}. All three approaches report improved completeness of quality requirements and aligned terminology among stakeholders. 

We summarize that knowledge management solutions are not well studied in the quality requirements context. We also note that there are similarities between knowledge management and quality models -- which is also evident as one of the studies used ISO 9126~\cite{ISO9126}. 

\subsubsection{Others}
We found one paper evaluating MOQARE, a misuse oriented approach to find adversarial quality requirements~\citeSR{SLR121}.  Another paper found that creating a clearer template and instructions for how to write a specification improved not only the quality requirements but also the attitude towards quality requirements~\citeSR{SLR289}. One paper evaluates a method for how to select an appropriate technique depending on the relevant quality requirements and context factors such as lifecycle phase, etc~\citeSR{SLR608}. Kopczy\'nska et al. experimented on a template approach for eliciting quality requirements~\citeSR{SLR575}. They define a template as a regular expression. The experiment is performed with students in the third year at university. They find that using templates improved completeness and overall quality of the quality requirements. However, using templates did not speed up the elicitation process. As this is a small scale validation, more research is needed to understand how it performs in a realistic setting.

\section{Discussion}\label{sec:discussion}

The results of our systematic literature review indicate that there are many quality requirements engineering aspects that warrant further research. The small number of studies found -- 84 papers over 30 years -- point to a lack of studies. Furthermore, it seems to us that there is a divide between academically proposed solutions and needs of practitioners. 

\subsection{RQ1 Which empirical methods are used to study quality requirements? }
A central research question when performing a systematic literature review is to try and answer a specific question through empirical evidence from several studies~\cite{Kitchenham07guidelinesfor}. There is a tendency towards more empirical studies on quality requirements in Figure~\ref{fig:type-year}. We found 1-5 papers per year in the 1990s and 4-11 papers per year in the 2010s. However, considering that the scientific community is producing more and more papers every year, the tendency to more empirical studies in quality requirements might be smaller than that of the empirical software engineering community as a whole. A naive search on Scopus for ``empirical software engineering" resulted in 50-100 papers per year during the 1990s and 450-800 papers per year in the 2010s. When we further limit the result to ``requirements", we end up with 1-20 in the 1990s and 100-250 papers per year in the 2010s. Ambreen et al. argue that quality requirements research is one of the emerging areas in their mapping study~\cite{ambreen2018empirical}. In their mapping study, they found 1-7 papers per year in the 1990s and 17-32 papers 2005-2012\footnote{Their results do not include papers after 2012. Hence, we choose a slightly different interval.}. In a recent paper in IEEE Software, quality requirements or non-functional requirements occurred in less than 1\% as a keyword in papers in the Requirements Engineering Journal and at the REFSQ conference and not at all in the top-ten list for the Requirements Engineering conference~\cite{tenbergen2019requirements}. This is an indication that the statement that quality requirements being an emerging area of research needs to be nuanced. We believe there is a need for further research on quality requirements. However, with the results we got from our study, it seems that the research on quality requirements might be less directed towards the practical challenges facing industry. This, however, is something that needs more research to be confirmed. 

It should be noted that we have not included studies where specific sub-characteristics of quality requirements, such as security or usability, are the focus. Ambreen et al., however, included also those~\cite{ambreen2018empirical}. Hence, the figures we presented might be on the lower side regarding the number of empirical studies. At the same time, we included 84 papers whereas Ambreen et al. found 36 papers. If we limit our results to papers before 2013 -- i.e. to the same period as Ambreen et al. -- we found 43 papers. 

In terms of the type of research performed, we see a similar distribution among validation, evaluation, and experience studies as Ambreen et al.~\cite{ambreen2018empirical} -- see Table~\ref{tab:method_overview}. Evaluation research is the most common and case study in industry the largest category. As noted in Section~\ref{sec:overall_results}, however, the rigor overall is weak. Furthermore, replications, longitudinal, and evaluation studies of a particular solution are rare. Similarly, we found only 4 experiments that can infer that the field could benefit from larger research initiatives planning several studies over many years. This would enable researchers to plan multiple studies, combining different research methods and larger sampling of the relevant population. 

\subsection{RQ2 What are the problems and challenges for quality requirements identified by empirical studies?}
We found studies claiming quality requirements are treated the same way as other types of requirements. We also found studies claiming companies do prioritize quality requirements and other studies claiming quality requirements are not handled properly. Furthermore, we found several studies attempting to discern which of the sub-characteristics -- such as security or performance -- are more important than others in a certain context. The studies we found are conducted in different contexts, domains, and different research methods. We interpret the empirical data that, on the one hand, different sub-characteristics might warrant individual attention, on the other, different research questions and methods are needed to understand and address industry-relevant challenges. We hypothesize that the requirements engineering community has not yet found a good way to analyze the quality requirements practices and challenges. We believe longitudinal studies is one way forward to deepen the understanding of quality requirements over the lifecycle of a product family rather than individual products or even just one point in time. This, of course, is not isolated to quality requirements and would entail changes to how projects are funded to allow for bigger projects. 

Overall, we summarize that quality requirements are written informally without a specific notation or modeling approach, there is no clear industry practice, and documentation of quality requirements seems to be performed in the same way as other requirements. Firstly, we speculate that there is a lack of understanding from practitioners of the specific needs for quality requirements, as they do not seem to prioritize separate handling. We believe the key to improving the understanding of the importance of quality requirements is to better understand the consequences and implications of quality requirements. Secondly, we believe the results imply that quality requirements engineering need to align well with other requirements topics as the cost of separate handling might deter usage. 

We did not find any studies connecting quality requirements to business value nor success criteria such as timely delivery or increased sales. There are some attempts to connect user satisfaction to quality requirements and defect flows to quality requirements. There are also some studies trying to discern perspectives internally -- e.g. that of architects. We also found some studies trying to understand app reviews as a source of requirements. However, we did not find any other studies attempting to identify other sources or ways of eliciting or analyzing new requirements. Interestingly, we found only one study explicitly on open-source software. There are some studies on agile methods, but we did not find any studies on DevOps, nor any studies bridging the gap between engineering and business. We propose to study quality requirements in more contexts, especially software ecosystems where commercial organizations cooperate both on development as well as operations. Furthermore, open-source is increasingly common. Hence, we believe these areas warrant more research. 

We believe -- at least for certain quality requirements sub-characteristics -- it is key to understand the actual usage by actual users to discern which quality requirements to address. We propose data-driven approaches as an important trend for quality requirements, seen with different automatic analysis techniques of app stores. Another example of unexploited potential is customer service data -- whether through, e.g., issues or reviews. We lack a clear data-driven perspective, where usage data, as one example, is studied with a quality requirements perspective. Furthermore, we believe the community needs to understand and address the connection between quality requirements and external factors such as business value or project success. An often-cited study by Finkelstein et al. claim quality requirements as one source of problem~\cite{finkelstein1996comedy}. We recognize, though, that this type of research is both challenging to design, expensive to perform, and difficult to get rigorous and relevant results. We see a need for future research in quality requirements to study not quality requirements in isolation but as part of larger studies where quality requirements are one of the research questions. 

\subsection{RQ3 Which quality requirements solution proposals have been empirically validated?} 
Quality models -- as an over-arching principle -- is the most common approach to elicitation. The different approaches typically propose tailoring of a generic quality model -- often ISO9126~\cite{ISO9126} or ISO25010~\cite{ISO25010} -- and a process with workshops to elicit and analyze quality requirements. The studies overwhelmingly report success with such strategies, albeit few have concrete and validated numbers for effort nor lead-time. 
However, it seems to us that there is no solid evidence of the cost-effectiveness of quality models, a lack of evidence of adherence over time, nor clear data for success factors from a more complete scope when using quality models. We see an opportunity for companies already using -- or willing to introduce -- quality models which should make it reasonable to conduct longitudinal studies on quality models as future work. 

We only found one modeling approach - goal modeling. Goal modeling is -- similar to quality models -- well researched. There are experiments and case studies. However, we could not find any surveys nor action research. Goal modeling is, just as quality models, integrated often integrated into a process encompassing all aspects of quality requirements. We interpret the lack of quality requirements modeling studies as follows: 1) Modeling of quality requirements cannot be separated from the modeling of other requirements. This is in line with quality requirements artifact studies, which often reports similar or identical specification for quality requirements and other requirements. 2) The modeling solutions proposed by academia are not something practitioners see as applicable or relevant. 

We found very few studies on data-driven requirements engineering~\cite{maalej2015toward} in the context of quality requirements. Rather, there seems to be a focus on the requirements specification, e.g. with quality models, goal modeling, and with the automatic analysis studies on mining specifications for quality requirements. Furthermore, we could not find any approaches integrating quality requirements engineering with, for example, DevOps and continuous experimentation. We see a gap for studies on how user feedback (reviews, customer service data, etc.) and usage data (measurements when using a software product or service) can be used for quality requirements prioritization, elicitation, and release planning. Furthermore, we propose that evaluating the trend over time as a means to better understand the connection between quality requirements and user satisfaction. Especially, we see an opportunity with the lead-time from an improvement in the quality requirements -- through the implementation -- and the lead-time from downward-trending user satisfaction to actions taken -- using various measurements -- are important to develop relevant early forecasting metrics and improved prioritization mechanisms which consider estimations of the user satisfaction. 

We note that the validation studies we found tend to report an improvement and rarely conclude that the proposals as a whole do not work. The explanations can be many, but publication bias can be one, another might be confirmation bias. We see this as an indication that empirical software engineering as an area is still developing and maturing.  

\subsection{Limitations and Threats to Validity}

\textbf{Construct validity}
Construct validity refers to the decisions on method and tools, and whether they are appropriate for the research questions. We utilized a hybrid search strategy. The risk is if the start set for the snowballing approach is insufficient, all papers will not be found when snowballing. However, Mourao et al. recently published an evaluation that a hybrid strategy is an appropriate alternative~\cite{MouraoHybrid2020}. Hence, the hybrid method is considered to be appropriate for our research questions. 

Start set I includes 274 papers between the years 1991 and 2014. Start set II includes 173 papers between 1990 and 2019 -- of which 70 are from the period 2015-2019. The snowballing iterations included 83 papers from 1976-2019. 447 of the 530 (84\%) of papers screened come from Start set I and the Start set II of papers. Furthermore, based on our experience, we believe we have included several key papers on empirical evidence on quality requirements. Hence, we believe that the fact that most of the included papers come from Start set I and the Start set II indicates that we have likely found the majority of all relevant papers. This implies that our method selection is appropriate. 

\textbf{Internal validity}
A validity threat is if papers are excluded even though they should have been included or erroneous classification. We ensured that all border-line cases were screened by at least two researchers and a sample of all papers was also screened by at least two researchers to mitigate this, see Section~\ref{sec:validity}. We also followed a pre-defined method thoroughly. In the end, the process resulted in: 

\begin{itemize}
    \item For Start set I, all papers were screened by at least 2 researchers. 
    \item For Start set II, 61\% of the papers were screened by at least two researchers. 
    \item For the two snowballing iterations, 29\% and 60\% of the papers respectively were screened by at least two researchers. 
    \item 66\% of the 193 papers included in the full read were read by at least two researchers, including reviewing the classification. 
    \item 83\% of excluded papers in the screening step and 77\% of the excluded papers in the full read step were read by at least two researchers. 
\end{itemize}

Excluded papers were reviewed more often by at least two researchers than included papers, mitigating the threat of excluding papers that should be included. We argue that threats to internal validity are low. 

There is a risk that we missed relevant papers are we excluded relevant papers as we excluded papers focusing on a specific type of quality requirements, e.g. performance or security. The risk is related to, on the one hand, terminology and, on the other, that the specific empirical study is on a specific type but the method or phenomena applies to quality requirements in general. For the former, we believe that our selection of search terms in the extended step is by far the most prevalent, hence, it should not be a big problem. Furthermore, since we also use a snowballing approach, this threat is further minimized. For the latter, we cannot completely dismiss the threat as some empirical evidence might not be presented on other papers even though the results might be applicable. We excluded 5 papers based on these criteria. Hence, we conclude that even though this is a threat, it is not likely to largely impact the internal from our paper.

\textbf{Conclusion validity}
We followed a systematic process to address threats to the conclusion validity. Furthermore, we report the steps and results in such a way that it should be possible to replicate them. The threat to conclusion is the inclusion/exclusion primarily, which entails a human judgment and thereby susceptible for errors. However, as mentioned for internal validity threats, we used a peer review process among the authors to minimize the threats of human errors.

\textbf{External validity}
External validity concerns the applicability of the results of our study. We believe the systematic hybrid process limits this threat as we do not exclude research communities nor do we exclude studies even if particular keywords are missing. However, we did not analyze different domains in detail as that information was not available in sufficient detail in enough studies. It might be that different domains exhibit different characteristics in terms of quality requirements engineering. Hence, the results should be applied after careful consideration.

\section{Conclusion}\label{sec:conclusion}
The results of our systematic literature review indicate that there are many quality requirements engineering aspects that warrant further research. We judge that 84 papers over 30 years point to a lack of studies. However, this is something that should be studied in more detail to be confirmed. Furthermore, it seems to us that there is a divide between academically proposed solutions accepted by practitioners. The proposed solutions are rarely evaluated in realistic settings -- and replications are non-existent. Furthermore, practitioners rarely report using any specific approach for quality requirements. A the same time, the existing surveys are small, have an unclear sample and population, and are rarely connected to any theory. We, therefore, hypothesize that overall, there is a lack of clear empirical evidence for what software developing organizations should adopt. This, again, is something that warrants further research to understand the needs of practitioners and their relation to proposed solutions found in the literature. 

For practitioners, there are some recommendations of what has worked in realistic contexts. Quality models with the associated processes, QUPER, and the NFR method have been reported as useful in several studies. However, it is not clear what the return of investment is nor the long-term effect. Still, we believe our results indicate those to be a good starting point if an organization should improve their quality requirements practices. Furthermore, goal modeling has been evaluated in academic settings with positive results. However, we could not find any evaluations in a realistic setting specifically for quality requirements. In the context where a document or specification is received, different automatic analysis approaches seem to be able to help in identifying quality requirements. However, we could not find any available tools nor clear integration in the overall software engineering process. Hence, even though these solutions show potential, the effort needed to apply them in practice is unclear. 

For researchers, we see a need for longitudinal studies on quality requirements. There are examples of solutions evaluated at one point in time. However, we could not find any studies on the long-term effect and costs of changing how companies work with quality requirements. We believe that the product or portfolio lifecycle is particularly under-researched. Furthermore, we believe there is a lack of understanding of the challenges and needs in realistic settings, as the solutions proposed by researchers seem to fail in getting acceptance from practitioners. This is a rather difficult issue for individuals to address, rather the requirements engineering community should try to establish a new way of performing research where larger and longer studies are viable. 

Furthermore, there are only a few studies on sources of quality requirements in general and data-driven alternatives specifically. We believe there is potential in sources such as usage data, customer service data, and continuous experimentation to complement stakeholder analysis, expert input, and focus groups. The former has the potential to take in a breadth of input closer to the actual users while the latter will focus on fewer persons' opinions or experiences which will be less representative of the actual usage. 

We limited our systematic literature review to quality requirements in general and excluding sub-categories such as security or usability. We believe it would be interesting to perform a similar study on the different sub-categories. For one, there might be differences in the sub-categories both regarding the strength of evidence and the types of solutions proposed. On the other, it might be that it does not make sense to have one solution for all types of quality requirements categories.

\bibliographystyle{splncs04}

\bibliography{references}

\bibliographystyleSR{splncs04}
\bibliographySR{sr}

\begin{table}[]
\caption{All case study papers, with context, scale, whether they are exploratory or studying a specific solution, and the main theme. }
\label{tab:casestudies_all}
\begin{tabular}{@{}llllcl@{}}
\toprule
Reference       & Year & Context     & Scale  & Exploratory & Theme \\ \midrule
\citeSR{SLR223} & 1999 & Academic    & Medium & N           & NFR method \\
\citeSR{SLR287} & 2001 & Academic    & Small  & N           & QUARCC, S-COST        \\
\citeSR{SLR284} & 2004 & Academic    & Small  & N           & NFR method  \\
\citeSR{SLR661} & 2014 & Academic    & Small  & N           & SeNOR, NoRT        \\
\citeSR{SLR42}  & 2015 & Academic    & Small  & Y           & Understanding QRs         \\ \midrule
\citeSR{SLR170} & 1995 & Industry    & Small  & N           & NFR method         \\
\citeSR{SLR289} & 1999 & Industry    & Large  & N           & Gilb style        \\
\citeSR{SLR291} & 1999 & Industry    & Large  & N           & Quality profile        \\
\citeSR{SLR293} & 2001 & Industry    & Medium & Y           & Metrics        \\
\citeSR{SLR224} & 2001 & Industry    & Medium & Y           & Process evaluation        \\
\citeSR{SLR263} & 2003 & Industry    & Small  & Y           & Process evaluation        \\
\citeSR{SLR285} & 2005 & Industry    & Medium & N           & IESE NFR Method        \\
\citeSR{SLR297} & 2007 & Industry    & Medium & N           & Quality model        \\
\citeSR{SLR296} & 2007 & Industry    & Large  & N           & QUPER        \\
\citeSR{SLR121} & 2008 & Industry    & Medium & N           & MOQARE        \\
\citeSR{SLR281} & 2008 & Industry    & Large  & N           & QUPER        \\
\citeSR{SLR340} & 2008 & Industry    & Large  & Y           & Understanding QRs        \\
\citeSR{SLR230} & 2009 & Industry    & Small  & N           & Light-weight process        \\
\citeSR{SLR252} & 2011 & Industry    & Small  & N           & Business process and i*         \\
\citeSR{SLR91}  & 2011 & Industry    & Small  & N           & QUPER        \\
\citeSR{SLR662} & 2012 & Industry    & Small  & N           & QUAMOCO        \\
\citeSR{SLR79}  & 2012 & Industry    & Large  & N           & QUPER        \\
\citeSR{SLR231} & 2012 & Industry    & Small  & Y           & Process evaluation        \\
\citeSR{SLR547} & 2012 & Industry    & Medium & N           & NFR evaluation model        \\
\citeSR{SLR608} & 2013 & Industry    & Small  & N           & QAT Framework        \\
\citeSR{SLR201} & 2013 & Industry    & Large  & Y           & Process evaluation        \\
\citeSR{SLR217} & 2013 & Industry    & Medium & Y           & Understanding QRs        \\
\citeSR{SLR197} & 2015 & Industry    & Large  & N           & QUPER        \\
\citeSR{SLR272} & 2015 & Industry    & Small  & N           & SQIMF        \\
\citeSR{SLR215} & 2015 & Industry    & Medium & Y           & Understanding QRs        \\
\citeSR{SLR36}  & 2016 & Industry    & Large  & Y           & Understanding QRs        \\
\citeSR{SLR267} & 2017 & Industry    & ?      & N           & MERLiNN        \\
\citeSR{SLR233} & 2017 & Industry    & Large  & N           & ADEG-NFR        \\
\citeSR{SLR237} & 2017 & Industry    & Medium & Y           & Sources        \\
\citeSR{SLR182} & 2017 & Industry    & Large  & Y           & Developers view        \\
\citeSR{SLR275} & 2018 & Industry    & Large  & Y           & Sources        \\
\citeSR{SLR667} & 2019 & Industry    & Large  & Y           & Understanding QRs        \\
\citeSR{SLR185} & 2019 & Industry    & Large  & Y           & Developers view        \\
\citeSR{SLR274} & 2019 & Industry    & Large  & Y           & Understanding QRs        \\
\citeSR{SLR279} & 2019 & Industry    & Large  & Y           & Process evaluation        \\ \midrule
\citeSR{SLR162} & 2001 & Mixed       & Medium & N           & NFR method         \\
\citeSR{SLR341} & 2007 & Mixed       & Small  & N           & QRF        \\ \midrule
\citeSR{SLR236} & 2010 & Open source & Medium & Y           & Understanding QRs        \\ \bottomrule
\end{tabular}
\end{table}

\end{document}